\documentclass[letterpaper]{article} 
\usepackage{aaai2026}  
\usepackage{times}  
\usepackage{helvet}  
\usepackage{courier}  
\usepackage[hyphens]{url}  
\usepackage{graphicx} 
\usepackage{natbib}  
\usepackage{caption} 
\usepackage{algorithm}
\usepackage{algorithmic}

\usepackage{booktabs}
\usepackage{multirow}
\usepackage{arydshln}
\usepackage{booktabs}
\usepackage{multirow}
\usepackage{array}
\usepackage{makecell}
\usepackage{dsfont}
\usepackage{amsmath}

\usepackage{amsfonts}
\usepackage[table]{xcolor}  
\usepackage{subcaption} 

\usepackage{newfloat}
\usepackage{listings}
\DeclareCaptionStyle{ruled}{labelfont=normalfont,labelsep=colon,strut=off} 
\floatstyle{ruled}
\newfloat{listing}{tb}{lst}{}
\floatname{listing}{Listing}
\newcommand{\MD}{\textsc{TDBA}}
\newcommand{\BT}{\textsc{BackTime}}
\newcommand{\pdelta}{\Delta^\mathtt{PTN}}
\newcommand{\gdelta}{\Delta^\mathtt{TGR}}

\title{Beyond Immediate Activation: Temporally Decoupled\\ Backdoor Attacks on Time Series Forecasting}
\author{
    Zhixin Liu\textsuperscript{\rm 1,2,3},
    Xuanlin Liu\textsuperscript{\rm 3},
    Sihan Xu\textsuperscript{\rm1,2,4}\thanks{Corresponding author},
    Yaqiong Qiao\textsuperscript{\rm 1,2,4},
    Ying Zhang\textsuperscript{\rm 3},
    Xiangrui Cai\textsuperscript{\rm 1,2,3}
}
\affiliations{
    \textsuperscript{\rm 1}Key Laboratory of Data and Intelligent System Security, Ministry of Education, China (DISSec)\\
    \textsuperscript{\rm 2}Tianjin Key Laboratory of Visual Computing and Intelligent Perception (VCIP)\\
    \textsuperscript{\rm 3}College of Computer Science, Nankai University, Tianjin, China\\
    \textsuperscript{\rm 4}College of  Cryptology and Cyber Science, Nankai University, Tianjin, China\\

    xusihan@nankai.edu.cn
    
}

\usepackage{bibentry}

\begin{document}

\maketitle

\begin{abstract}

Existing backdoor attacks on multivariate time series (MTS) forecasting enforce strict temporal and dimensional coupling between triggers and target patterns, requiring synchronous activation at fixed positions across variables. However, realistic scenarios often demand delayed and variable-specific activation.
We identify this critical unmet need and propose \MD{}, a temporally decoupled backdoor attack framework for MTS forecasting. By injecting triggers that encode the expected location of the target pattern, \MD{} enables the activation of the target pattern at any positions within the forecasted data, with the activation position flexibly varying across different variable dimensions.
\MD{} introduces two core modules: (1) a position-guided trigger generation mechanism that leverages smoothed Gaussian priors to generate triggers that are position-related to the predefined target pattern;  and (2) a position-aware optimization module that assigns soft weights based on trigger completeness, pattern coverage, and temporal offset, facilitating targeted and stealthy attack optimization.
Extensive experiments on real-world datasets show that TDBA consistently outperforms existing baselines in effectiveness while maintaining good stealthiness.
Ablation studies confirm the controllability and robustness of its design.

\end{abstract}

\begin{links}
    \link{Code}{https://github.com/steven705/TDBA}
\end{links}

\section{Introduction}

Time series forecasting is fundamental to many real-world applications, including finance~\cite{finance1, finance2}, climate modeling~\cite{climate1, climate2}, epidemic prediction~\cite{disease2}, and traffic management~\cite{ traffic2}. 
To improve forecasting accuracy, a variety of deep learning models have been developed, including MLP-based methods~\cite{zeng2022DLinear, wang2024TimeMixer}, diffusion-based models~\cite{yuan2024diffusionts, li2025diffusionbased}, Transformer architectures~\cite{zhou2021informer, wu2022autoformer}, and more recently, time-series adaptations of large language models (LLMs)~\cite{jin2024timellm, liu2025timecma}.

\begin{figure}[htbp]
    \centering
    \begin{subfigure}[b]{0.48\columnwidth} 
        \centering
        \includegraphics[width=\textwidth]{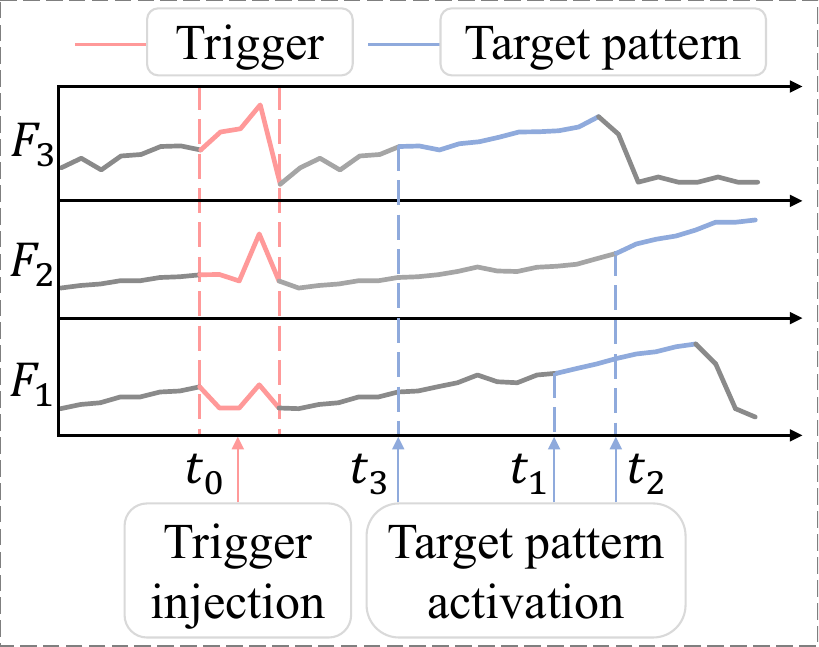} 
        \caption{}
    \label{fig:trigger-theory}
    \end{subfigure}
    \hfill 
    \begin{subfigure}[b]{0.48\columnwidth} 
        \centering
        \includegraphics[width=\textwidth]{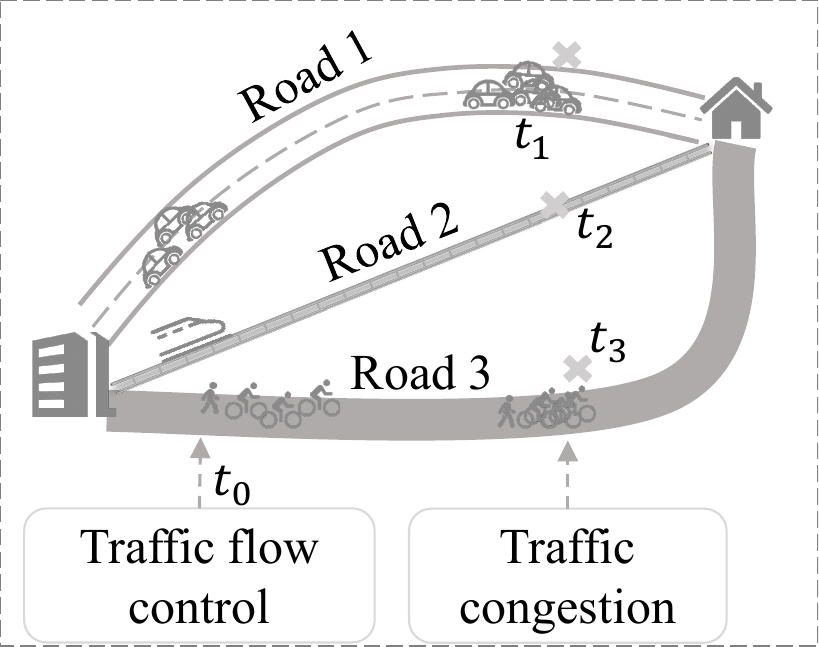} 
        \caption{}
    \label{fig:real-world}
    \end{subfigure}
   \caption{
    (a) Theoretical trigger injections at $t_0$ activating cone-shaped target patterns at $t_1$, $t_2$, and $t_3$ across $F_1$, $F_2$, and $F_3$. (b) Corresponding real-world traffic scenario where control at $t_0$ induces congestion on three roads. $F_1$–$F_3$ in (a) correspond to flow sensors on roads 1–3 in (b).
    }
    \label{fig:attack-overview}
\end{figure}

Recent studies have demonstrated that MTS models are vulnerable to backdoor attacks, where attackers can implant malicious behaviors through data poisoning during the training phase~\cite{ding2022towards,huang2025revisiting, lin2024backtime}. BackTime~\cite{lin2024backtime} was the first to formalize the backdoor threat in the field of MTS forecasting. 
Its attack mechanism involves injecting triggers into historical input data and implanting predefined target patterns into future output data. After training, the model learns the association between the presence of triggers and the activation of target patterns in predictions. 
However, \BT{} imposes rigid alignment constraints: the target pattern must appear immediately after the trigger and at identical positions across all variable dimensions. Such temporal and spatial synchronization significantly limits its flexibility and stealthiness, especially in real-world scenarios that exhibit delayed or asynchronous responses across variables.

We identify this as a critical yet unmet challenge in multivariate time series backdoor attacks. Taking the traffic flow forecasting scenario as an example, the attacker's core objective is to create phased traffic disruptions by manipulating prediction models. Triggers can be injected through diverse means: either via physical intervention such as artificially controlling traffic flow at current intersections, or direct manipulation of the forecasting model that is generating false congestion predictions to mislead the public into believing specific road segments are congested, inducing detours to alternative routes and ultimately causing systemic traffic disorder.

Specifically, the main highway linking suburban and downtown areas typically experiences peak traffic between 17:00 and 19:00. The attacker aims to inject triggers around 18:00, causing malicious predictions to activate one hour later,coinciding with real congestion and thereby intensifying it. For auxiliary roads, which serve as primary detours, predictions are activated 30 minutes earlier to block escape routes in advance. Meanwhile, roads surrounding metro hubs require predictions to be delayed by 2.5 hours to exploit post-rush transit surges and maximize disruption. Similar demands arise in other domains: financial scenarios may require predictions to indicate stock price increases occurring several days later, while energy systems need delayed triggers to mislead peak electricity load forecasts.

In the aforementioned scenario, the congestion situation corresponds to the term target pattern in backdoor attacks.
Specifically, the manifestation of congestion in time series data is a continuous increase starting from a certain moment; and causing such a congestion situation to occur is referred to as the activation of the target pattern. The malicious behavior of an attacker at the current moment is called trigger injection, which is specifically manifested as the manipulation and modification of historical data.

To achieve delayed and asynchronous activation of target patterns across different dimensions,
we propose a novel framework, Temporally Decoupled Backdoor Attack, termed \MD{}.
The key technical challenge lies in enabling the trigger generator to accurately learn the positional information of the target pattern within the forecast window, while supporting dimension-specific and temporally misaligned activations. 
\MD{} has two key modules: the first is a \textbf{position-guided trigger generation} module, which encodes the expected activation position of the target pattern through a smoothed Gaussian distribution, providing a differentiable position supervision signal for trigger generation. This enables the trigger generator to perceive the position information of the target pattern and further adapt to the delay requirements of different variables; the second is a \textbf{position-aware optimization} module, which extends the soft identification mechanism in the sliding window scenario, incorporates the offset between the target pattern and the trigger into weight calculation, and designs a new loss function to dynamically focus on effective backdoor activation samples. This further enhances the trigger's perception of the target pattern and reduces deviations in non-attacked regions by means of the optimized loss function.
The expected effects and application scenarios are illustrated in Figure~\ref{fig:attack-overview}.

In summary, our main contributions are as follows.
\begin{itemize}
    \item We identify a critical unmet need in backdoor attacks on MTS forecasting: the lack of support for delayed and asynchronous activation of target patterns across different variables under existing frameworks.
    
    \item We propose \MD{}, the first temporally decoupled backdoor attack framework to address this gap. It enables target patterns to be activated at attacker-specified delayed positions across different dimensions, and supports activation at any locations within the forecasted data. This is achieved through two core modules: a position-guided trigger generation mechanism and a position-aware optimization module with a dedicated loss function.
    
    \item Extensive experiments on five real-world datasets validate the superiority of \MD{}. It achieves precise control over target pattern positions (with the lowest attacked-position error $M_p^a$) while maintaining high stealthiness (with comparable or lower unaffected-position error $M_p^c$) compared to baselines. Ablation studies further confirm the necessity of each core component.
\end{itemize}

\section{Related Work}

\paragraph{Multivariate Time Series Forecasting.}

Recent years have witnessed rapid advances in MTS forecasting, driven by the growing availability of sequential data and the demand for accurate prediction. Beyond traditional statistical methods, deep learning approaches have shown superior performance in modeling complex temporal dependencies. In particular, Transformer-based models~\cite{zhou2021informer, wu2022autoformer, zhou2022fedformer, liu2024itransformer} and diffusion-based forecasting frameworks~\cite{yuan2024diffusionts, li2025diffusionbased} have achieved state-of-the-art results across various benchmarks. Additionally, the emergence of time-series-specific large language models (LLMs)~\cite{jin2024timellm, liu2025timecma, shi2025timemoe} further expands the modeling capabilities in this domain.

\paragraph{Backdoor Attacks in Deep Learning.}

Backdoor attacks are a class of security threats where models are trained to behave normally on clean inputs but output malicious predictions when a specific trigger is present~\cite{gu2017badnets, wang2019defsi, zhao2020clean, liu2020reflection}. These attacks can be broadly classified into two categories: poisoning-based methods that manipulate the training data~\cite{gu2017badnets, li2020rethinking, sarkar2021facehack}, and training-stage attacks that alter the model optimization process~\cite{doan2021lira, ding2022towards, jiang2023backdoor}.

While most early efforts focused on vision and NLP tasks, backdoor attacks on time series have recently gained attention~\cite{ding2022towards,huang2025revisiting,lin2024backtime,jiang2023backdoor,dong2025trojantime}. TimeTrojan~\cite{ding2022towards} was the first to introduce targeted backdoor attacks in time series classification. It injects imperceptible perturbations into the input to trigger misclassification at inference time. Follow-up studies explored more advanced mechanisms, such as frequency-domain perturbations~\cite{huang2025revisiting}, to exploit the spectral vulnerabilities of temporal models.

\section{Notations and Preliminaries}

\subsection{Multivariate Time Series Forecasting}

A MTS data consists of temporal observations over multiple correlated variables. Formally, the entire data is denoted as \( \mathbf{X} \in \mathbb{R}^{T \times N} \), where \( T \) represents the total number of timesteps, and \( N \) denotes the number of variables observed at each timestep. Specifically, let \( \mathbf{X} = \{x_1, x_2, \dots, x_N\} \), where each \( x_i \in \mathbb{R}^{T} \) denotes the time series of the \( i \)-th variable.

In MTS forecasting tasks, a sliding window mechanism~\cite{zhou2021informer,chen2021autoformer,wu2022timesnet} is commonly adopted to construct input–output pairs for model training. Let \( h \) and \( f \) denote the input and output window lengths, respectively. At any reference timestamp \( t_i \), the historical data is denoted as
\(
X_{t_i, h} = \mathbf{X}[t_i - h : t_i, :] \in \mathbb{R}^{h \times N},
\)
and the corresponding future data is defined as
\(
X_{t_i, f} = \mathbf{X}[t_i : t_i + f, :] \in \mathbb{R}^{f \times N}.
\)
For consistency, we refer to \( X_{t_i, h} \) and \( X_{t_i, f} \) as the historical data and future data throughout the paper.

\subsection{Threat Model and Attack Scenario}
\BT{} introduces backdoor attacks in the context of MTS forecasting by constructing a poisoned training set. Once a downstream model is trained on this dataset, it will produce  malicious outputs when specific triggers are embedded in the input.

The downstream model, denoted as $f_d$, refers to any multivariate time series forecasting model that may be deployed in real-world applications and is trained on the poisoned dataset constructed by the attacker.
This includes proprietary or open-source models commonly used for tasks such as traffic forecasting, power consumption prediction, and financial analysis.

We consider a black-box threat scenario, where the attacker has full access to the training dataset but no knowledge of the downstream model’s architecture~\cite{ding2023black,lin2024backtime}, optimization algorithm, or hyperparameter configurations. 
This setting captures practical situations in which machine learning practitioners or organizations train private models using publicly available or externally provided datasets that may have been poisoned beforehand.

Under this threat scenario, the attacker aims to achieve two goals:
\begin{itemize}
    \item \textbf{Forecasting Integrity on Clean Inputs:} The downstream model should maintain high forecasting performance on clean (non-poisoned) inputs, ensuring its behavior appears normal and indistinguishable from a benign model.
    \item \textbf{Controlled Manipulation on Triggered Inputs:} Upon the injection of a trigger into the historical data, the model is expected to output a predefined target pattern at specific positions within the forecast window. This pattern should be restricted to the attacker-specified variables and timesteps, while maintaining accurate predictions for all remaining dimensions.
\end{itemize}
Our proposed attack framework is designed based on this threat scenario.

\section{Temporally Decoupled Backdoor Attack}

This section provides an overview of the proposed \emph{Temporally Decoupled Backdoor Attack Framework}, which enables temporal decoupling between the trigger and the target pattern in MTS forecasting, allowing the target pattern to be activated at any positions within the forecast horizon.
Section~\ref{sec:4.1} defines the temporal injection strategy for both the trigger and the target pattern, establishing the foundation of our attack framework. 
Building on this, Section~\ref{sec:4.2} and Section~\ref{sec:4.3} respectively describe how to guide the trigger generator under any target pattern positions in the forecasted data, and how to optimize it effectively. 

\subsection{Temporally Decoupled Injection Strategy}\label{sec:4.1}

Backdoor attacks in MTS forecasting first involve injecting a trigger into the historical data and injecting a predefined target pattern in the future data.
In our framework, both operations are performed directly on the clean training set
\( \mathbf{X}_{\text{train}} \in \mathbb{R}^{T \times N} \).

Following \BT{}, we first select a subset of timestamps \( \mathcal{T}^{\mathtt{ATK}} \subset \{1, \dots, T\} \) that exhibit high prediction error under a surrogate forecasting model \( f_s \). This surrogate model is also a MTS forecasting model, but its parameters are independent of those of the downstream model \( f_d \).

For each selected timestamp \( t_i \in \mathcal{T}^{\mathtt{ATK}} \), we inject a trigger \( \mathbf{g} \in \mathbb{R}^{t^{\mathtt{TGR}} \times |\mathcal{S}|} \) into the historical data over a set of attacked variables \( \mathcal{S} \subset \{1, \dots, N\} \). The injection is performed dimension-wise as:
\begin{equation}
\small
    \mathbf{X}[t_i - t^{\mathtt{TGR}} : t_i,\ s] 
    \leftarrow 
    \mathbf{X}[t_i - t^{\mathtt{TGR}} - 1 ,\ s] \oplus \mathbf{g}[:, s], 
    \quad \forall s \in \mathcal{S},
\end{equation}
where \( \oplus \) denotes element-wise additive perturbation, and \( t^{\mathtt{TGR}} \) is the temporal length of the trigger.

Simultaneously, a predefined target pattern \( \mathbf{p} \in \mathbb{R}^{t^{\mathtt{PTN}} \times |\mathcal{S}|} \) is injected into the future data, where \( t^{\mathtt{PTN}} \) denotes the length of the target pattern. To allow temporal and dimension-wise flexibility, we assign each variable \( s \in \mathcal{S} \) an individual offset \( \Delta t_i^{(s)} \in \mathbb{N} \), forming an offset set:
\(
\mathbf{T}_i = \{ \Delta t_i^{(s)} \}_{s \in \mathcal{S}}.
\)
The target pattern is then asynchronously injected across variables as:
\begin{equation}
    \mathbf{X}[t_i + \Delta t_i^{(s)} : t_i + \Delta t_i^{(s)} + t^{\mathtt{PTN}},\ s] \leftarrow \mathbf{p}[:, s], \quad \forall s \in \mathcal{S},
\end{equation}
where each \( \Delta t_i^{(s)} \in \mathbb{N} \) denotes the relative temporal offset between the end of the trigger and the intended start position of the target pattern for variable \( s \).
To preserve stealthiness, the amplitude of the injected perturbations is constrained:
\begin{equation}
    \| \mathbf{g}_{:,s} \|_{\infty} \leq \gdelta_s, \quad 
    \| \mathbf{p}_{:,s} \|_{\infty} \leq \pdelta_s, \quad \forall s \in \mathcal{S},
\end{equation}
where \( \gdelta_s \) and \( \pdelta_s \) denote dimension-specific perturbation budgets, each defined in proportion to the standard deviation of the corresponding variable \( s \).

We refer to \( \widetilde{X}_{t_i, h} \) as the poisoned historical data, and \( \widetilde{X}_{t_i, f} \) as the manipulated future data that embeds the predefined target pattern \( \mathbf{p} \) at the designated positions. 
The above injection operations collectively form the original poisoned training set \( \widetilde{\mathbf{X}}_{\text{train}} \), which will undergo further optimization in subsequent steps.

\subsection{Position-guided Trigger Injection}\label{sec:4.2}

\paragraph{Position Guidance Matrix.}
To guide the trigger generator to attend to the positional information of the target pattern within the future data when generating triggers, we construct a \emph{position guidance matrix} \( \mathbf{A}_d \in \mathbb{R}^{f \times |\mathcal{S}|} \) based on the offset set \( \mathbf{T}_i = \{\Delta t_i^{(s)}\}_{s \in \mathcal{S}} \).

For each attacked variable \( s \in \mathcal{S} \), the corresponding guidance vector \( \mathbf{A}_d[:, s] \) is constructed using a Gaussian kernel~\cite{scholkopf2002learning} centered at \( \Delta t_i^{(s)} \), defined as:
\begin{equation}
\small
    \mathbf{A}_d[d, s] = \frac{1}{Z} \exp\left( -\frac{(d - \Delta t_i^{(s)})^2}{2\sigma^2} \right),
    \label{eq:guass}
\end{equation}
where \( d \in \{0, 1, \dots, f - t^{\mathtt{PTN}} \} \), \( \sigma \) controls the spread of the distribution, and \( Z \) is a normalization constant ensuring \( \sum_d \mathbf{A}_d[d, s] = 1 \).

This smoothed positional encoding provides a differentiable, dimension-aware supervisory signal to the trigger generator, encouraging it to align trigger characteristics with the desired activation positions of the target pattern. Notably, this design allows the position guidance matrix \( \mathbf{A}_d \) to be integrated seamlessly into different types of trigger generator architectures.
In this work, we instantiate this mechanism using two designs described below.

\paragraph{GCN-based Trigger Generator with Position Guidance.}

The GCN-based trigger generator, adapted from \BT{}~\cite{lin2024backtime}, synthesizes adaptive triggers based on structural correlations among variables. It constructs a correlation graph \( \mathbf{A} \in \mathbb{R}^{|\mathcal{S}| \times |\mathcal{S}|} \) using frequency-filtered features from each variable’s global time series and applies a GCN to generate the perturbations.

Formally, given the historical data \( \mathbf{X}_{t_i, t^{\mathtt{BEF}}} \in \mathbb{R}^{t^{\mathtt{BEF}} \times |\mathcal{S}|} \) before trigger, the trigger is computed as:
\begin{equation}
    \hat{\mathbf{g}}_{t_i} = \mathbf{A} \cdot \mathbf{X}_{t_i, t^{\mathtt{BEF}}}^{\top} \cdot \mathbf{W},
\end{equation}
where \( \mathbf{W} \in \mathbb{R}^{t^{\mathtt{BEF}} \times t^{\mathtt{TGR}}} \) is a learnable projection matrix.

To incorporate positional priors, we inject the guidance matrix \( \mathbf{A}_d \) as an auxiliary term:

\begin{equation}
    \hat{\mathbf{g}}_{t_i} = \mathbf{A} \cdot \mathbf{X}_{t_i, t^{\mathtt{BEF}}}^{\top} \cdot \mathbf{W} 
    + \mathbf{A} \cdot \mathbf{A}_d^{\top} \cdot \mathbf{W}_d,
\end{equation}
where \( \mathbf{W}_d \in \mathbb{R}^{f \times t^{\mathtt{TGR}}} \) enables the generator to align trigger generation with the desired target activation position.

We modify the original output scaling by replacing the $tanh(\cdot)$ function with the $softsign(\cdot)$ function to achieve smoother output values and better stealthiness:
\begin{equation}
    \mathbf{g}_{t_i} = \Delta^{\mathtt{TGR}} \cdot \text{softsign}(\hat{\mathbf{g}}_{t_i}).
\end{equation}

\paragraph{Inverse Forecasting Trigger Generator with Position Guidance.}

The inverse forecasting-based trigger generator (InverseTgr) generates triggers by reversing the standard forecasting direction. Instead of generating future predictions from historical data, it generates historical trigger based on a reversed forecast sequence that embeds the predefined target pattern.

Formally, given the manipulated forecast sequence \( \widetilde{\mathbf{X}}_{t_i, f} \) containing the injected target pattern, InverseTgr takes it as input along with positional guidance \( \mathbf{A}_d \) and temporal marker embeddings \( \mathbf{X}_{\text{mark}} \), and outputs the corresponding trigger:
\begin{equation}
    \mathbf{g}_{t_i} = f_{\mathtt{Inv}}\left( \widetilde{\mathbf{X}}_{t_i, f},\ \mathbf{X}_{\text{mark}},\ \mathbf{A}_d \right),
\end{equation}
where \( f_{\mathtt{Inv}}(\cdot) \) denotes the inverse forecasting model.

The guidance matrix \( \mathbf{A}_d \) is integrated into the generator as a soft positional prior, encouraging the model to focus on forecast regions where the target pattern is most prominent.

\subsection{Position-aware Backdoor Optimization}\label{sec:4.3}
\paragraph{Position-aware Soft Identification.}
MTS forecasting models are typically trained using a sliding window approach.This setup implies that, during training on poisoned data, individual sliding windows may only partially include the injected trigger or the target pattern.A window may contain the full trigger in the input data while only partially covering the target pattern in the future data.

In such cases, it becomes challenging to determine whether a sample should be involved in backdoor optimization (e.g., whether the input window fully includes the trigger), and how much optimization weight it should carry (e.g., how much of the target pattern is visible in the output).

To address this issue, we extend the soft identification mechanism originally proposed in \BT{} by introducing a position-aware weighting scheme that takes into account the following aspects: (1) whether the full trigger is included in the input window, (2) the proportion of the target pattern visible in the output, and (3) the offset between the trigger and the target pattern, which influences attack strength and visibility.

Specifically, we define the soft identification weight for each timestamp \( t_i \) as:
\begin{equation}
    \beta(t_i) = 
    \mathds{1}(c^{\mathtt{TGR}}_{t_i} = t^{\mathtt{TGR}}) \cdot
    \eta\left(\frac{\mathcal{P}[t_i, \Delta t]}{t^{\mathtt{PTN}}}\right) 
    \cdot \phi(\Delta t).
    \label{eq:pos_soft_identification}
\end{equation}
This formulation includes three key components. The \textbf{trigger completeness indicator} \( \mathds{1}(c^{\mathtt{TGR}}_{t_i} = t^{\mathtt{TGR}}) \) is a binary variable that determines whether the current input window contains all \( t^{\mathtt{TGR}} \) steps of the trigger, where \( c^{\mathtt{TGR}}_{t_i} \) denotes the number of trigger timesteps actually observed in the historical input window at timestamp \( t_i \). Only samples with fully observed triggers are eligible to contribute to backdoor optimization. The \textbf{pattern coverage term} \( \eta\left( \frac{\mathcal{P}[t_i, \Delta t]}{t^{\mathtt{PTN}}} \right) \) quantifies the fraction of the target pattern that appears within the forecast window, where \( \mathcal{P}[t_i, \Delta t] \) denotes the number of visible target pattern steps starting from offset \( \Delta t \). We adopt a linear form \( \eta(x) = x \), which increases the weight proportionally with the visible portion of the target pattern. The \textbf{offset penalty} \( \phi(\Delta t) = \exp(-\lambda \Delta t) \) discourages injections that occur too far into the forecast window. This exponential decay mitigates the impact of delayed activations and favors target pattern placements closer to the beginning of the prediction horizon.


\paragraph{Loss function based on Position-aware Soft Identification.}
After generating the \( \widetilde{\mathbf{X}}_{\text{train}} \) using the trigger generator, we still adopt the \( f_s \) mentioned in Section~\ref{sec:4.1} to simulate the training process of the downstream model. The gradients from \( f_s \) are then utilized to optimize the trigger generator in a gradient-based manner.

To ensure the optimization remains effective under the sliding-window setting, we incorporate the previously proposed position-aware soft identification function \( \beta(t_i) \), which selectively weighs each poisoned sample based on its relevance to the backdoor objective.

Specifically, we decompose the prediction error into two components:  
(1) \( \mathcal{L}_{\mathtt{tp}} \), the loss computed over positions and variables where the target pattern is injected, which reflects attack effectiveness;  
(2) \( \mathcal{L}_{\mathtt{cln}} \), the loss computed over the remaining clean regions, which encourages prediction consistency and suppresses side effects.

The position-aware attack loss is defined as:
\begin{equation} \label{eq:atk_loss_pos_main}
    \mathcal{L}_{\mathtt{atk}} = 
    \sum_{t_i = t}^{t + K}
    \beta(t_i) \cdot 
    \mathcal{L}_{\mathtt{tp}}(t_i) + 
    \lambda_{\mathtt{cln}} \cdot 
    \mathcal{L}_{\mathtt{cln}}(t_i), 
    \quad \forall t \in \mathcal{T}^{\mathtt{ATK}},
\end{equation}
where \( K \) denotes the sliding range over which poisoned segments \([ \widetilde{X}_{t_i, h}, \widetilde{X}_{t_i, f} ]\) are sampled, and \( \lambda_{\mathtt{cln}} \) controls the trade-off between attack effectiveness and stealthiness.

The two loss components are formally defined as:
\begin{equation}
    \mathcal{L}_{\mathtt{tp}}(t_i) =
    \frac{1}{|\mathcal{M}_{\mathtt{tp}}|} \sum_{J \in \mathcal{M}_{\mathtt{tp}}}
    \left | f_s(\widetilde{X}_{t_i, h})[J] - \widetilde{X}_{t_i, f}[J] \right|^2,
\end{equation}
\begin{equation}
    \mathcal{L}_{\mathtt{cln}}(t_i) =
    \frac{1}{|\mathcal{M}_{\mathtt{cln}}|} \sum_{J \in \mathcal{M}_{\mathtt{cln}}}
    \left| f_s(\widetilde{X}_{t_i, h})[J] - \widetilde{X}_{t_i, f}[J] \right|^2,
\end{equation}
where  \( \mathcal{M} \) is formally an index set of the form \( (t,n) \), representing a specific  timestamp and variable index.
\( \mathcal{M}_{\mathtt{tp}} \) and \( \mathcal{M}_{\mathtt{cln}} \) denote the subsets of forecasted positions corresponding to injected and clean regions, respectively.

To further regularize the shape and amplitude of generated triggers, we add an \( L_2 \) penalty term:
\begin{equation}
    \mathcal{L}_{\mathtt{reg}} = 
    \lambda_{\mathtt{reg}} \cdot \| g \|_2^2,
\end{equation}
where \( \lambda_{\mathtt{reg}} \) is a regularization coefficient.

The final objective for training the trigger generator is:
\begin{equation} \label{eq:G_loss}
    \mathcal{L}_{G} = \mathcal{L}_{\mathtt{atk}} + \mathcal{L}_{\mathtt{reg}}.
\end{equation}

\subsection{Training  Strategy }
Our proposed \MD{} framework proceeds as follows: 
First, target timestamps $\mathcal{T}^{\mathtt{ATK}} $ and corresponding positional offsets $\mathbf{T}_i$ are selected, and the position guidance matrix $\mathbf{A}_d$ is constructed.Then, the injection of the target pattern is accomplished.
Then, the trigger generator uses $\mathbf{A}_d$ and auxiliary inputs to produce triggers injected into the historical data, creating an initial poisoned dataset $\widetilde{\mathbf{X}}_{\text{train}}$.
A surrogate forecasting model $f_s$ is trained on $\widetilde{\mathbf{X}}_{\text{train}}$, with the trigger generator optimized iteratively via loss~\ref{eq:G_loss}.
Finally, the optimized trigger generator is applied to all target timestamps $\mathcal{T}^{\mathtt{ATK}} $ and $\mathbf{T}_i$ to generate the final $\widetilde{\mathbf{X}}_{\text{train}}$.
The algorithm workflow is presented in Appendix A.

\section{Experiments}

\begin{table*}[t]
\small  
\centering
\setlength{\tabcolsep}{1.1mm}  
\begin{tabular}{c|c|ccc|ccc|ccc|ccc|ccc}
\toprule
\multirow{2}{*}{Dataset} & \multirow{2}{*}{Model} 
& \multicolumn{3}{c|}{Random} 
& \multicolumn{3}{c|}{Manhattan} 
& \multicolumn{3}{c|}{BackTime} 
& \multicolumn{3}{c|}{TDBA-Inv} 
& \multicolumn{3}{c}{TDBA-Gcn} \\
&  
& $M_c$ & $M_p^c$ & $M_p^a$
& $M_c$ & $M_p^c$ & $M_p^a$
& $M_c$ & $M_p^c$ & $M_p^a$
& $M_c$ & $M_p^c$ & $M_p^a$
& $M_c$ & $M_p^c$ & $M_p^a$ \\
\midrule

\multirow{4}{*}{Weather} 
& TimesNet & 20.60 & 19.58 & 11.95 & 23.56 & 22.10 & 10.71 & 14.28 & 14.12 & 30.89 & 26.91 & 20.10 & \textbf{5.14} & 12.93 & 14.85 & 5.34 \\
& FEDformer & 9.68 & 9.99 & 12.92 & 9.27 & 8.30 & 5.95 & 9.39 & 10.24 & 21.34 & 10.72 & 12.65 & 3.09 & 11.19 & 9.96 & \textbf{3.06} \\
& Autoformer & 8.86 & 10.48 & 15.39 & 8.31 & 7.34 & 5.50 & 8.39 & 9.81 & 17.42 & 9.27 & 10.64 & \textbf{2.20} & 8.31 & 9.96 & 2.64 \\
& \cellcolor{gray!20}Average 
& \cellcolor{gray!20}13.05 & \cellcolor{gray!20}13.35 & \cellcolor{gray!20}13.42 
& \cellcolor{gray!20}13.71 & \cellcolor{gray!20}12.58 & \cellcolor{gray!20}7.39 
& \cellcolor{gray!20}10.69 & \cellcolor{gray!20}11.39 & \cellcolor{gray!20}23.22 
& \cellcolor{gray!20}15.63 & \cellcolor{gray!20}14.46& \cellcolor{gray!20}\textbf{3.48} 
& \cellcolor{gray!20}10.81 & \cellcolor{gray!20}11.59 & \cellcolor{gray!20}3.68 \\

\midrule

\multirow{4}{*}{PEMS03} 
& TimesNet    & 19.73 & 16.43 & 22.82 & 20.17 & 16.15 & 20.90 & 20.07 & 20.48 & 27.82 & 19.61 & 19.51 & 22.59 & 20.09 & 20.14 & \textbf{18.63} \\
& FEDformer   & 16.52 & 14.62 & 21.16 & 16.63 & 13.99 & 17.27 & 16.28 & 17.77 & 20.02 & 16.70 & 16.76 & 18.31 & 17.66 & 17.84 & \textbf{17.06} \\
& Autoformer  & 18.04 & 16.02 & 25.79 & 16.67 & 14.05 & 17.65 & 16.24 & 17.22 & 18.28 & 16.22 & 16.25 & 18.25 & 17.21 & 17.40 & \textbf{16.94} \\
& \cellcolor{gray!20}Average 
& \cellcolor{gray!20}18.10 & \cellcolor{gray!20}15.69 & \cellcolor{gray!20}23.26 
& \cellcolor{gray!20}17.82 & \cellcolor{gray!20}14.73 & \cellcolor{gray!20}18.61 
& \cellcolor{gray!20}17.53 & \cellcolor{gray!20}18.49 & \cellcolor{gray!20}22.04 
& \cellcolor{gray!20}17.51 & \cellcolor{gray!20}17.51 & \cellcolor{gray!20}19.72
& \cellcolor{gray!20}18.32 & \cellcolor{gray!20}18.46 & \cellcolor{gray!20}\textbf{17.21} \\
\midrule

\multirow{1}{*}{PEMS04} 
& \cellcolor{gray!20}Average 
& \cellcolor{gray!20}23.00 & \cellcolor{gray!20}19.94 & \cellcolor{gray!20}32.04 
& \cellcolor{gray!20}22.43 & \cellcolor{gray!20}19.01 & \cellcolor{gray!20}38.54 
& \cellcolor{gray!20}22.28 & \cellcolor{gray!20}22.96 & \cellcolor{gray!20}30.75 
& \cellcolor{gray!20}22.42 & \cellcolor{gray!20}22.42 & \cellcolor{gray!20}\textbf{27.40}
& \cellcolor{gray!20}21.88 & \cellcolor{gray!20}21.88 & \cellcolor{gray!20}27.85\\
\midrule

\multirow{1}{*}{PEMS08} 
& \cellcolor{gray!20}Average 
& \cellcolor{gray!20}19.73 & \cellcolor{gray!20}16.98 & \cellcolor{gray!20}27.33 
& \cellcolor{gray!20}19.72 & \cellcolor{gray!20}16.12 & \cellcolor{gray!20}32.58 
& \cellcolor{gray!20}18.99 & \cellcolor{gray!20}19.58 & \cellcolor{gray!20}25.10 
& \cellcolor{gray!20}19.21 & \cellcolor{gray!20}19.37 & \cellcolor{gray!20}17.79 
& \cellcolor{gray!20}18.32 & \cellcolor{gray!20}18.46 & \cellcolor{gray!20}\textbf{17.52} \\
\midrule

\multirow{1}{*}{ETTh1} 
& \cellcolor{gray!20}Average 
& \cellcolor{gray!20}1.97 & \cellcolor{gray!20}1.54 & \cellcolor{gray!20}3.73 
& \cellcolor{gray!20}1.92 & \cellcolor{gray!20}1.48 & \cellcolor{gray!20}3.51 
& \cellcolor{gray!20}1.90 & \cellcolor{gray!20}2.00 & \cellcolor{gray!20}2.74 
& \cellcolor{gray!20}2.07 & \cellcolor{gray!20}1.93 & \cellcolor{gray!20}\textbf{2.22} 
& \cellcolor{gray!20}2.00 & \cellcolor{gray!20}2.28 & \cellcolor{gray!20}2.90 \\
\bottomrule
\end{tabular}
\caption{
    Performance comparison of different backdoor attack strategies in terms of $M_c$, $M_p^c$, and $M_p^a$. 
    The target pattern shape is set to cone.
    TDBA-Inv refers to  \MD{} based on InverseTgr, and TDBA-Gcn refers to  \MD{} based on TgrGCN
    Lower $M_p^a$ indicates stronger attack effectiveness, while lower $M_p^c$ suggests better stealthiness. The best results for each row (lowest $M_p^a$) are highlighted in \textbf{bold}.
    Please refer to Appendix C.1 for full results.
}
\label{tab:main_ret}
\end{table*}

\subsection{Experimental Setup}
\paragraph{Datasets.}
We evaluate our method on the five real-world MTS datasets: 
Weather~\cite{lin2024backtime}, PEMS03~\cite{song2020spatial}, PEMS04~\cite{song2020spatial}, PEMS08~\cite{song2020spatial}, and ETTh1~\cite{zhou2021informer}.
For the PEMS dataset, we utilize only the traffic flow features collected by different sensors for training and evaluation.
The datasets are split into training, validation, and testing sets using a ratio of 6:2:2. Detailed dataset statistics and descriptions are provided in Appendix B.1.


\paragraph{Metrics.}
To comprehensively evaluate the effectiveness and stealthiness of backdoor attacks on time series forecasting models, we adopt the following three metrics.
$M_c$ represents the Mean Absolute Error (MAE) of the forecasted output when the model is given clean (non-poisoned) historical inputs. This reflects the model's general forecasting performance under normal conditions.
$M_p^a$ denotes the MAE of the forecasted output when the historical input contains an injected trigger. The error is computed only over the positions and variables where the target pattern is present in the output. This metric measures the attack’s effectiveness by assessing how closely the model's predictions align with the predefined malicious pattern.
$M_p^c$ refers to the MAE of the forecasted output under the same poisoned input as above, but calculated only over the positions and variables that are not affected by the target pattern. This captures the stealthiness of the attack, with lower values indicating less collateral deviation from clean predictions.
Lower values of $M_c$ and $M_p^c$ indicate better forecasting performance and stealthiness, respectively, while a lower $M_p^a$ implies higher attack success.

\paragraph{Baselines.}
We compare our method against the following representative baselines: 
\BT{}, Random, and Manhattan.
\BT{} trains multiple trigger generators, each corresponding to a specific positional offset \( \Delta t \). For each training run, a fixed offset \( \Delta t \) is uniformly applied across all attack timestamps and all attacked variables, resulting in the same target pattern position throughout. After training separate models for all possible offsets, final results are reported by averaging the performance over all values of \( \Delta t \). While effective, this setting restricts the flexibility of target injection and introduces high training overhead.
Random adopts the same poisoned timestamp selection strategy as \BT{}, but replaces learned triggers with fixed random perturbations. Specifically, triggers are sampled from a uniform distribution over the range \( [-\gdelta, \gdelta] \), and reused across different injection positions and samples.
Manhattan also follows the poisoning strategy of \BT{} for timestamp selection. It constructs triggers by identifying clean input segments from the training set whose future trajectories exhibit the lowest Manhattan distance to a predefined target pattern. The historical data preceding these matched segments are then used as surrogate triggers.

At test time, all baselines are evaluated on a fixed poisoned test set to ensure fair comparison.
For each time step in the sliding window of the test set, we independently generate a random offset set \( \mathbf{T}_i \) ,and this process is performed only once.

\begin{table}[t]
\centering
\setlength{\tabcolsep}{1.2mm}
\begin{tabular}{c|ccc|ccc}
\toprule
\multirow{2}{*}{Methods}
& \multicolumn{3}{c|}{Upward trend} 
& \multicolumn{3}{c}{Up and down}  \\
& $M_c$ & $M_p^c$ & $M_p^a$
& $M_c$ & $M_p^c$ & $M_p^a$ \\
\midrule
Random         & 18.15 & 15.96 & 24.18  & 17.93 & 25.53 & 15.73 \\
Manhattan      & 16.64 & \textbf{13.92} & 17.73  & 16.54 & \textbf{13.96} & 18.14 \\
BackTime       & 16.22 & 16.97& 20.04  & 16.22 & 17.49& 19.14  \\
TDBA-Inv  & 16.35 & 16.37 & 23.33 & 16.23 & 16.26 & 19.00 \\
TDBA-GCN  & \textbf{15.84} & 15.83 & \textbf{17.70} & \textbf{15.04}  & 15.83 & \textbf{15.50} \\
\bottomrule
\end{tabular}
\caption{
Performance on the PEMS03 dataset using Autoformer under different target pattern shapes.
The best results for each row (lowest $M_c$,lowest $M_p^c$,lowest $M_p^a$) are highlighted in \textbf{bold}.
}
\label{tab:main_ret_2}
\end{table}

\paragraph{Experiment Protocol.}
We follow the experimental settings of \BT{} to ensure fair comparisons. 
Specifically, we randomly select $\alpha_t = 3\%$ of all available timestamps as poisoned injection points, and for each poisoned sample, we inject the target pattern into $\alpha_s = 30\%$ of the variable dimensions. 
The attacked variable dimensions are randomly selected for each injection instance.
Each sample is processed using a sliding window of 24 timesteps, consisting of 12 input steps and 12 output steps. 
The length of the trigger is fixed at \( t^{\mathtt{TGR}} = 4 \), while the length of the target pattern is set to \( t^{\mathtt{PTN}} = 7 \). 
Throughout all experiments, we evaluate three representative target pattern styles: \emph{cone}, \emph{upward trend}, and \emph{up-and-down}. 
All parameter settings above are aligned with \BT{} to ensure consistency and fair evaluation across methods.
The value of $\sigma$ in Equation~\ref{eq:guass} is set to 1 throughout all experiments.
Details of the hyperparameter settings and descriptions of the predefined target pattern are provided in Appendix B.2.

We evaluate the attack performance on three widely used forecasting models: TimesNet~\cite{wu2022timesnet}, FEDformer~\cite{zhou2022fedformer}, and Autoformer~\cite{chen2021autoformer}, which serve as downstream models in a black-box setting.
The specific designs of these three models follow the implementation in the \BT{} framework.

Among them, FEDformer is adopted as the surrogate model $f_s$. 
Each experiment is run three times, and the average performance is reported. 

\begin{figure}[h]
    \centering
    \includegraphics[width=0.95\columnwidth]{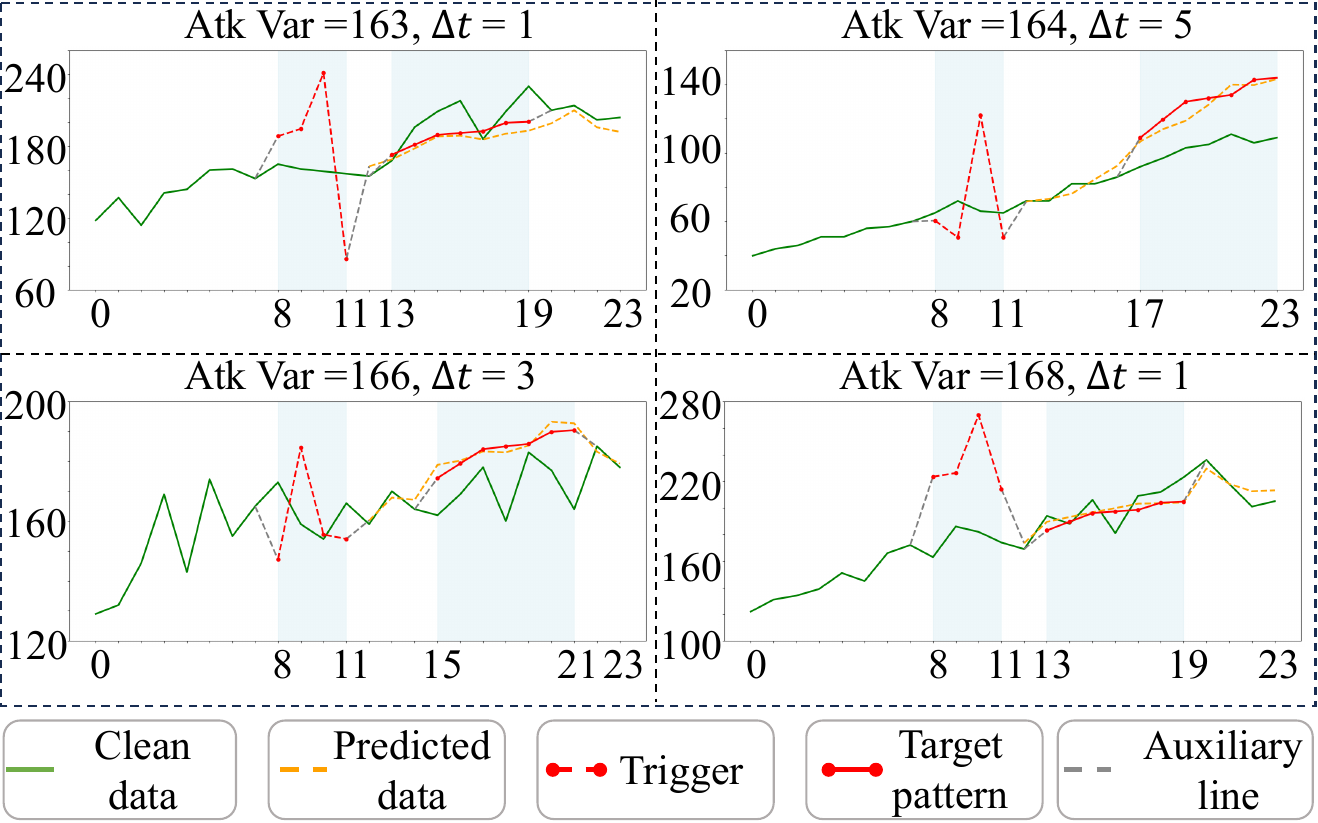}
    \caption{
    Visualization of the \MD{} on the PEMS04 dataset using Autoformer. Four dimensions (163, 164, 166, 168) are attacked, each with its own assigned positional offset. The auxiliary lines in the figure captions are used to connect the trigger or target pattern to the clean data at both ends, maintaining visual uniformity.}
    \label{fig:vis}
\end{figure}

\subsection{Experimental Results}
\paragraph{Quantitative Results.}
Our main experimental results are presented in Table~\ref{tab:main_ret}, where the target pattern shape is set to \emph{cone}. 
As shown in Table~\ref{tab:main_ret},  \textsc{TDBA-Inv} and \textsc{TDBA-Gcn}, demonstrate superior attack performance across all evaluated datasets. Both methods consistently achieve the lowest $M_p^a$ in almost all model-dataset combinations, significantly outperforming existing baselines and indicating stronger capabilities in manipulating predictions at targeted timestamps.
Specifically, \textsc{TDBA-Gcn} obtains the best $M_p^a$ on two out of five datasets, while \textsc{TDBA-Inv} slightly outperforms on other datasets.
For example, on the Weather dataset, \textsc{TDBA-Inv} reduces the average $M_p^a$ to 3.48, which is considerably lower than 23.22 achieved by \BT{}, highlighting its precise temporal attack effectiveness. 
In terms of stealthiness, compared to \BT{}, our methods maintain comparable or even lower $M_c$ and $M_p^c$, indicating that the injected trigger have minimal impact on the non-targeted timesteps and dimensions in the forecasted data. 

Table~\ref{tab:main_ret_2} compares the \MD{} against baseline approaches when the target patterns are in the shapes of an upward trend and an up-and-down trend. The experimental results show that our method achieves the best \( M_p^a \) and \( M_c \) values under both shapes, demonstrating the effectiveness and stealthiness of our attack.

\paragraph{Qualitative Analysis.} 
As illustrated in Figure~\ref{fig:vis}, the model successfully outputs the predefined target pattern at the specified forecast positions, despite varying offsets across dimensions. Notably, the predictions for unaffected regions remain highly accurate, demonstrating the strong stealthiness of our attack.

\begin{table}[t]
\centering
\begin{tabular}{c|cc}
\toprule
Method Variant & $M_p^c$ & $M_p^a$ \\
\midrule
\multicolumn{3}{c}{TDBA-Inv} \\
\midrule
Full Model & 16.25 & 18.25 \\
A1         & 18.84 (+15.9\%) & 35.34(+93.64\%) \\
A2         & 19.82 (+21.9\%) & 29.48(+61.53\%) \\
\midrule
\multicolumn{3}{c}{TDBA-Gcn} \\
\midrule
Full Model & 17.40 & 16.94 \\
A1         & 26.44 (+51.9\%) & 30.64 (+81.0\%) \\
A2         & 21.44 (+23.2\%) & 28.99 (+71.1\%) \\
\bottomrule
\end{tabular}
\caption{
Ablation study on key components of \MD{}. 
The values in parentheses indicate the percentage increase in $M_p^c$ and $M_p^a$ relative to the Full Model, reflecting degradation in stealthiness and attack precision.
Please refer to Appendix C.2 for full results on all dataset.
}
\label{tab:Ablation}
\end{table}

\subsection{Model Analysis}

\paragraph{Ablation Analysis.}  
To assess the contributions of key components in our framework, we conduct an ablation study on two representative variants, summarized in Table~\ref{tab:Ablation}:

\begin{itemize}
    \item \textbf{A1 (w/o Position Guidance Matrix)}: This variant removes the positional guidance matrix \( \mathbf{A}_d \), so the trigger generator loses access to the positional structure of the injected target pattern.

    \item \textbf{A2 (w/o Position-aware Optimization Objective)}: This variant discards the proposed position-aware backdoor loss. Instead, it formulates the attack as a standard regression problem: given the poisoned input window \( \widetilde{X}_{t_i,h} \), it directly optimizes the output toward the target pattern \( \widetilde{X}_{t_i,f} \) using a  MAE loss.
\end{itemize}

Table~\ref{tab:Ablation} shows the performance comparison with the full model. After removing the positional guidance matrix (A1), both the $M_p^a$ and $M_p^c$ increase significantly, indicating a substantial degradation in attack effectiveness and stealthiness due to loss of positional information.
When without the position-aware optimization objective (A2), the $M_p^c$ rises substantially. This observation suggests that that the triggers cause more interference on clean positions and variables, reflecting reduced stealth.
These results demonstrate the importance of both Positional Guidance Matrix and the Position-aware Optimization Objective in improving the effectiveness and stealthiness of our approach.

\paragraph{Stealthiness Assessment.} 
To evaluate the imperceptibility of the poisoned samples generated by \MD{}, we employ a representative unsupervised anomaly detection method, USAD~\cite{audibert2020usad}, to detect potential anomalies that might reveal the presence of backdoor triggers. For each dataset, the anomaly detector is first trained on the clean test set to learn normal temporal patterns. It is then applied to the poisoned training set $ \widetilde{\mathbf{X}}_{\text{train}} $, where timestamps with injected triggers are treated as anomalies. The area under the ROC curve (AUC-ROC) is computed to measure the detector's ability to identify manipulated timestamps.

As shown in Table~\ref{tab:Anomaly Detection}, the AUC-ROC scores across all tested datasets are consistently close to 0.5. This indicates that the anomaly detector performs no better than random guessing, thereby validating the high stealthiness of our approach.

\begin{table}[t]
\centering
\begin{tabular}{c|c|c|c}
\toprule
\multirow{1}{*}{Method}
& \multicolumn{1}{c|}{PEMS03}
& \multicolumn{1}{c|}{PEMS04} 
& \multicolumn{1}{c}{PEMS08}   \\
\midrule
BackTime          & 0.5279  & 0.5747 & 0.5389  \\
TDBA-Inv      & 0.5275  & 0.4949  & 0.5163  \\
TDBA-GCN      & 0.5313     & 0.4961    & 0.4977  \\
\bottomrule
\end{tabular}
\caption{
Anomaly detection results using the USAD algorithm on datasets generated by our method (TDBA-Inv, TDBA-GCN) and the BackTime method, with the tested datasets being PEMS03, PEMS04, and PEMS08.
}
\label{tab:Anomaly Detection}
\end{table}

\section{Conclusion}
In this work, we address a critical limitation in existing MTS forecasting backdoor attacks: their inability to support delayed and asynchronous activation of target patterns across different variables.
To bridge this gap, we propose \MD{}.
By introducing variable-specific positional offsets, \MD{} enables flexible and asynchronous target pattern activation in predicted data. 
It combines a position-guided trigger generation mechanism with a position-aware optimization objective. Experiments on five real-world datasets show that \MD{} achieves superior attack effectiveness compared to existing methods.  
Current limitations include support for single target pattern training and limited cross-domain transferability.
Future work will explore multi target pattern learning and domain-adaptive trigger generation.

\clearpage
\section*{Acknowledgments}
This work was supported by the National Natural Science Foundation of China (Nos. 62572260, 62202245, and 62272163). We also gratefully acknowledge the support of Enowa Network Technology Co., Ltd., and thank Feipeng Dou and Hao Li for their insightful guidance.

\bibliography{aaai2026}

\end{document}